\documentclass[twocolumn,amsmath,amssymb,floatfix]{revtex4}

\usepackage{latexsym,amsmath,amssymb,theorem,epsfig}
\usepackage{graphicx}
\usepackage{subfigure}

\usepackage{fullpage}
\usepackage{amsmath}
\usepackage{amssymb}
\usepackage{bbm}
\usepackage{color}
\usepackage{ulem}

\usepackage{graphicx}
\usepackage{graphicx,subfigure}
\usepackage{epstopdf}
\usepackage[body={17.5cm, 21cm},right=2cm]{geometry}
\usepackage{amssymb}
\usepackage{amsmath}
\usepackage{psfrag}
\usepackage{epsfig}
\usepackage{cancel}

\newcommand{\kk}{{\bf k}}

\newcommand{\be}{\begin{equation}}
\newcommand{\ee}{\end{equation}}

\def\grad{\vec{\nabla}}
\def\gt{\rightarrow}

\def\z{\zeta}
\def\t{\tau}

\def\e{\epsilon}

\def\D{\Delta}
\def\h{\abs{h^{-1}}}

\def\ta{\abs{\tau}}

\def\dtau{\mathrm{d} \tau \,}
\def\dcx{\mathrm{d}^3 x \,}

\providecommand{\abs}[1]{\left\lvert#1\right\rvert}
\providecommand{\lsim}{\lesssim}
\providecommand{\gsim}{\gtrsim}

\begin{document}

\title{Towards a Cosmological Dual to Inflation}
\author{Justin Khoury and Godfrey E. J. Miller}
\affiliation{Center for Particle Cosmology, Department of Physics and Astronomy, University of Pennsylvania, 209 South 33rd Street, Philadelphia, PA 19104}

\begin{abstract}
We derive all single-field cosmologies with unit sound speed that generate scale invariant curvature perturbations on a dynamical attractor background. We identify three distinct phases: slow-roll inflation; a slowly contracting adiabatic ekpyrotic phase, described by a rapidly-varying equation of state; and a novel adiabatic ekpyrotic phase on a slowly expanding background. All of these yield identical power spectra. The degeneracy is broken at the 3-point level: unlike the nearly gaussian spectrum of slow-roll inflation, adiabatic ekpyrosis predicts large non-gaussianities on small scales. 
\end{abstract}

\maketitle

The observational evidence for primordial density perturbations with nearly scale invariant and gaussian statistics is compatible with the simplest inflationary scenarios. But is inflation unique? Are there {\it dual} cosmologies with indistinguishable predictions? Such questions are critical to our understanding of the very early universe.

Inflation not only generates scale invariant and gaussian density perturbations, it does so on an {\it attractor} background.
On super-horizon scales, the curvature perturbation on comoving hypersurfaces~\cite{zeta,separateuniverse}, denoted by $\zeta$,
measures differences in the expansion history of distant Hubble patches~\cite{separateuniverse}.
In single-field inflation, $\zeta$ approaches a constant at long wavelengths.  In the strict $k\rightarrow 0$ limit, $\zeta \gt \delta a/a$, so the perturbation simply renormalizes the scale factor of the background solution;  such a perturbation can be removed by an appropriate rescaling of global coordinates. For finite $k$, the perturbation cannot be completely removed, but different Hubble patches experience the same cosmological evolution, up to a shift of local time coordinates and a rescaling of local spatial coordinates. See~\cite{weinbergzeta} for a detailed discussion.

Achieving both scale invariance and dynamical attraction in alternative scenarios has proven challenging.
The $\zeta$ equation of a contracting, matter-dominated universe is identical to that of inflation~\cite{dust}, but $\zeta$ grows outside the horizon, indicating an unstable background. The contracting phase in the original ekpyrotic scenario~\cite{oldek1,oldek2,oldek3,oldek4,oldek5,oldek6}, with $V(\phi) = -V_0e^{-\phi/M}$, is an attractor~\cite{gratton,paoloniczal}, but the resulting spectrum is strongly blue~\cite{robertek,gratton,paoloniczal}. A scale invariant spectrum can be obtained through entropy perturbations~\cite{newek,2fieldek}, as in the New Ekpyrotic scenario~\cite{newek}, but this requires two scalar fields.

The {\it adiabatic ekpyrotic} mechanism~\cite{adiabaticek,linde,adiabaticek2,austin,Kinney:2010qa} proposed recently offers a counterexample: a single-field model for which
the background is a dynamical attractor and generates a scale invariant $\zeta$. The mechanism obtains for fairly simple potentials, such as $V(\phi) = V_0 (1-e^{-\phi/M})$ with $V_0 > 0$ and $M\ll M_{\rm Pl}$. Scale invariant perturbations are generated during the transition when $\epsilon \equiv -\dot{H}/H^2 = 3(1+w)/2$ rises rapidly from $\epsilon\ll 1$, where the constant term dominates, to $\epsilon \approx M_{\rm Pl}^2/2M^2\gg 1$, where the negative exponential term dominates. 

Another counterexample proposed recently relies on a rapidly-varying, superluminal sound speed $c_s(\tau)$~\cite{csm,csk,csfedo}.
See~\cite{csearlier,piazza} for earlier work. Even though the background is non-inflationary, $\zeta$ is amplified because the
sound horizon is shrinking. The growing mode is $\zeta \gt {\rm constant}$, and the resulting 2-point function is scale invariant.

The key lesson of these results is that relaxing some of the standard assumptions, such as
$w, c_s \approx {\rm constant}$, opens up new possibilities for generating perturbations.

In this paper, we derive the most general single-field cosmologies with $c_s =1$ that: $i)$ yield a scale invariant power spectrum
for $\zeta$; and $ii)$ are dynamical attractors, in the sense that $\zeta\rightarrow {\rm constant}$ is the
growing mode solution.  These conditions imply a second-order differential equation for $a(\t)$ whose exact solutions we classify.

The question of uniqueness is more than academic. If the Planck mission corroborates the predictions of the simplest single-field inflationary models, namely scale invariance and gaussianity, then the onus will be on theorists to determine whether inflation is unique in making these predictions. This work is an important first step in answering this critical and timely question.

We find only {\it three} possibilities: inflation, with $a(\tau)\sim 1/|\tau|$ and $\epsilon\approx {\rm constant}$;
the adiabatic ekpyrotic phase~\cite{adiabaticek,adiabaticek2}, with $\e \sim 1/\t^2$ on a slowly contracting background; 
and a novel adiabatic ekpyrotic phase on a background that first slowly expands, then slowly contracts~\cite{austin}.
At the 2-point level, therefore, the adiabatic ekpyrotic phases are dual to inflation. The degeneracy is broken at the 3-point level: adiabatic ekpyrosis generically predicts strongly scale dependent non-gaussianities, which limits the range of scale invariant modes that can be generated within the perturbative regime~\cite{adiabaticek2}.
Thus, if Planck finds no deviations from gaussianity, our work will imply that any alternative theory must either invoke multiple degrees of freedom or use an altogether different mechanism to generate density perturbations. 

Any portion of these phases can be used to devise novel early-universe models. Such scenarios
should explain the observed flatness and homogeneity, either through inflation or through an
ekpyrotic phase with $\epsilon \gg 1$~\cite{oldek1,Erick}. Moreover, a reheating mechanism must be specified.  In cases where
the universe is contracting, the Null Energy Condition must be violated to bounce
to an expanding phase, for instance within 4d effective theories~\cite{ghostNEC}.

For the purposes of this paper, however, we are solely interested in identifying {\it all} cosmological
phases that generate, with a single degree of freedom, super-horizon perturbations compatible with observations --- the
candidate duals to inflation. The idea of cosmological duals is not new~\cite{dust,gratton,boyle}, but we focus here on $\zeta$ instead of the Newtonian
potential~\cite{gratton,boyle} and specialize to attractor solutions by demanding that $\zeta\rightarrow {\rm constant}$.

\section{Set-up}
Our starting point is the quadratic action for $\z$, assuming $c_s =1$:
\be
S = M_{\rm Pl}^2 \int \dtau \dcx z^2 \left\{ ( \z' )^2 - ( \grad \z )^2 \right\},
\ee
where $z \equiv a \sqrt{2 \e}$ and primes denote derivatives with respect to conformal time $\tau$.  This yields the mode function equation for the canonically-normalized variable $v = z\,\zeta$:
\be
v_k''+\left(k^2-\frac{z''}{z}\right) v_k = 0\,,
\label{veom}
\ee
where $k$ is the comoving wavenumber. To generate a scale invariant spectrum from
adiabatic initial conditions, it is sufficient for $z$ to satisfy
\be
\frac{z''}{z} = \frac{2}{\tau^2}\,.
\label{z''}
\ee
Indeed, the solution to~(\ref{veom}) in this case is
\be
v_k = \frac{1}{\sqrt{2 k}M_{\rm Pl}} e^{-i k \t} \left(1-\frac{i}{k \t}\right)\,,
\label{vk}
\ee
which implies that $k^{3/2} |\zeta_k| = \sqrt{ 1 + k^2 \tau^2 }/ \sqrt{2} M_{\rm Pl} z |\tau|$.  As $\tau\rightarrow 0$, $k^{3/2} \abs{\z_k}$ is independent of $k$, as desired.

In addition to generating a scale invariant $\z_k$, our background must be a dynamical attractor.
Since $\z_k \sim 1/z |\tau|$ as $k\rightarrow 0$, the desired solution to~(\ref{z''}) is
\be
z \equiv \frac{\sqrt{2}}{m \ta}\,,
\label{zed}
\ee
where $m$ is an arbitrary scale. Combining~\eqref{vk} and~\eqref{zed} yields
$k^{3/2} \abs{\z_k} = m\sqrt{1 + k^2 \t^2}/2M_{\rm Pl}$, which is both scale invariant as $\t \gt 0$ and constant as $k \gt 0$. The observed amplitude
of $\zeta\sim 10^{-5}$ fixes $m\sim 10^{-5}M_{\rm Pl}$.

We pause to note that in an inflationary context the freeze-out or {\it $\z$-horizon} $\ta$ is usually identified with the {\it comoving Hubble horizon},
$h^{-1} \equiv 1/aH = a/a'$, but that more generally ({\it e.g.}, when $\e$ varies rapidly) the Hubble horizon and the $\z$-horizon can differ greatly.

Using the definition $z = a \sqrt{2 \e}$,~(\ref{zed}) implies
\be
\e = \frac{1}{a^{2} m^{2} \t^{2}}\,.
\label{ep}
\ee
Moreover, we can rewrite $\e  = -\dot{H}/H^2 =  {\rm d}H^{-1}/{\rm d}t$ in terms of the comoving Hubble horizon $h^{-1} = 1/aH$ as
\be
\left( h^{-1} + \t \right)' = \e\,.
\label{hub}
\ee
Combining~\eqref{ep} and~\eqref{hub} then gives a second-order differential equation for $a(\tau)$. Instead,
we will cast these as a pair of coupled first-order equations. By differentiating~\eqref{ep},
\be
(\log \sqrt{\e})' = - \t^{-1} - h\,. 
\label{boxep}
\ee
Once we specify the signs of $h$ and $\t$,~\eqref{hub} and~\eqref{boxep} become coupled ODEs
for $\h$ and $\e$. The behavior of~\eqref{boxep} will depend strongly on the relative magnitude of the Hubble horizon $\h$ and the $\z$-horizon $\ta$.  We will therefore say that the Hubble horizon is {\it inside the $\z$-horizon} when $\h < \ta$, and {\it outside the $\z$-horizon} when $\h > \ta$.

To solve these coupled equations, $h_{\rm fid}$ and $\epsilon_{\rm fid}$ must be specified at some fiducial time $\tau_{\rm fid} < 0$.   
To obtain a solution for $a(\tau)$, we can set $a_{\rm fid} = 1$ by a spatial rescaling $a\rightarrow \lambda a$, $\tau \rightarrow \tau/\lambda$.  The equation of state is of course
invariant, so $\epsilon_{\rm fid}$ fixes $\tau_{\rm fid}$ through~(\ref{ep}).
In practice, we will specify not $|h_{\rm fid}^{-1}|$ but the ratio $|h_{\rm fid}^{-1}|/|\tau_{\rm fid}|$, which is invariant under the above spatial rescaling.

\begin{figure}
\centering
\includegraphics[width=0.4\textwidth]{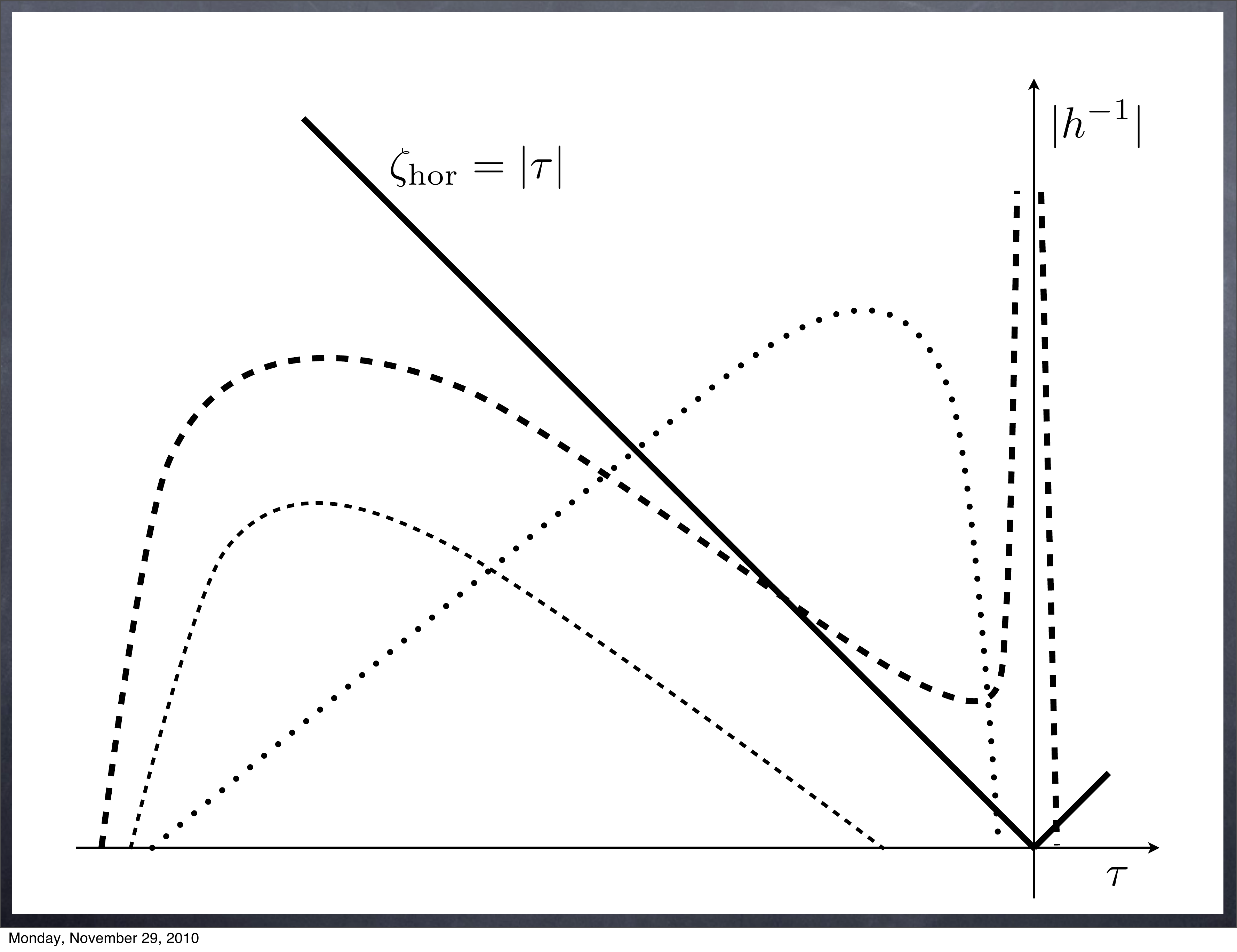}
\caption{Sketch of $|h^{-1}|$ for the contracting (dotted), expanding (dashed) and apex (thick dashed) branches of solutions.}
\label{sketchsolns}
\end{figure}

\section{Solutions}

While it is straightforward enough to integrate~\eqref{hub} and~\eqref{boxep} numerically, as we have done, to guide our intuition we also provide
a series of simple, analytical arguments that explain the general features of the solutions. By varying over all possible initial conditions, we find three families of solutions,
each of which is indexed by a single parameter and has finite duration, $\t_{\rm i} < \t < \t_{\rm f}$. See Fig.~\ref{sketchsolns} for a sketch of the solutions.
\subsection{Contracting Branch}
This case obtains if the universe is assumed contracting ($h_{\rm fid} < 0$)
at some fiducial time $\t_{\rm fid}< 0$. Then, as long as $h < 0$ and $\t < 0$,~(\ref{boxep}) implies $(\log \sqrt{\epsilon})' > 1/|\tau|$, hence $\epsilon$ increases
monotonically. Meanwhile,~(\ref{hub}) reduces to $\h' = 1- \e$,
thus $\h$ increases whenever $\e<1$, and decreases whenever $\e>1$. 
In fact, the bound from~(\ref{boxep}) implies that $\epsilon$ must pass through $\epsilon =1$,
at which point $\h$ hits a global maximum. A global maximum is a good point to specify a solution, so we denote the fiducial time in this case
as $\tau_{\rm fid}\rightarrow \tau_{\rm max} \equiv -m^{-1}$, where we set $a_{\rm max} = 1$ and $\e_{\rm max} = 1$.
All contracting solutions can therefore be indexed by the single parameter 
\be
 c \equiv \frac{|h^{-1}_{\rm max}|}{|\t_{\rm max}|} = m |h^{-1}_{\rm max}| > 0\,.
\label{cdef}
\ee

Before $\t_{\rm max}$, $a > 1$, so $\e < 1/m^2 \t^2$; after $\t_{\rm max}$, $a<1$, so $\e > 1/m^2 \t^2$. Integrating~(\ref{hub}) therefore yields
\be
m \h \leq f(\tau)  \leq c\,, 
\label{hconup}
\ee
where $f(\tau) \equiv c + 2 - m \ta - m^{-1} \ta^{-1}$, with the inequalities saturated at $\t_{\rm max}$. 
Since $m \h \leq c$, this implies that $h$ cannot change sign for $\t < 0$. Moreover, since $f(\tau)$ vanishes at $m\tau_\pm = -(c+2 \pm \sqrt{c}\sqrt{c+4})/2$, $h$ must diverge
at finite $\t$ in both the past and the future of $\t_{\rm max}$. Denoting the time of past and future divergences by $\tau_{\rm i}$ and $\tau_{\rm f}$, respectively,~(\ref{hconup})
implies $\t_+ < \t_{\rm i} < \t_{\rm max} < \t_{\rm f} < \t_- < 0$. Over the interval $\tau_{\rm i} \leq \tau \leq \tau_{\rm f}$, $a(\tau)$ contracts from $\infty$ to $0$, so $\tau_{\rm f}$ marks a big crunch singularity; from~(\ref{ep}), we conclude that $\e$ grows monotonically from $0$ to $\infty$.

The range of modes thus generated spans a factor of $k_{\rm max}/k_{\rm min} = |\tau_{\rm i}|/|\tau_{\rm f}| < |\tau_+|/|\tau_-|$.
From the definition of $\tau_{\pm}$ above, we have $|\tau_+|/|\tau_-|  =  (m \t_+)^2 < (c + 2)^2$, hence large
values of $c$ are required to generate a sufficiently broad range of scale invariant modes. From~(\ref{cdef}), this means
that $|h^{-1}|$ must venture far outside the $\zeta$-horizon, as sketched by the dotted line in Fig.~\ref{sketchsolns}. In this regime, $\e \approx 1/ m^2 \t^2$ and $a \approx 1$, which
is recognized as the {\it adiabatic ekpyrotic} phase proposed recently in~\cite{adiabaticek}. 

Nearly all scale invariant modes are produced while $|h^{-1}|$ is outside the $\zeta$-horizon. Integrating~(\ref{hub})
assuming $\e \approx 1/ m^2 \t^2$ gives $m |h^{-1}| \approx f(\tau)$, or
\be
m |h^{-1}| \approx c + 2 - m \ta - m^{-1} \ta^{-1}. \label{conhub}
\ee
For large $c$, horizon-equality ($|h^{-1}| = |\tau|$) occurs at
\be
\tau_{{\rm eq} +} \approx -\frac{c}{2 m}, \qquad \tau_{{\rm eq} -} \approx -\frac{1}{m c}, \label{concutoff}
\ee
hence this phase generates
$N_{\rm ek} = \log|\tau_{{\rm eq}+}|/|\tau_{{\rm eq}-}|  \approx 2\log c$ e-folds of modes.

Because $|h^{-1}|$ is outside the $\zeta$-horizon during mode production, perturbations freeze out
while {\it inside} the Hubble horizon and eventually exit Hubble by $\tau_{{\rm eq}-}$, when $|h^{-1}|$
re-enters the $\zeta$-horizon. If a finite portion of this solution is used in a broader scenario,
then some other dynamics must push these modes outside Hubble while maintaining
scale invariance. In~\cite{adiabaticek}, this is achieved through an ekpyrotic scaling phase with $\epsilon \approx c^2/2\gg 1$.

\subsection{Expanding Branch}

Suppose instead that the universe is expanding ($h_{\rm fid} > 0$) at some fiducial time $\tau_{\rm fid} < 0$.  It is helpful to rewrite~(\ref{hub}) and~(\ref{boxep}) in terms of the gap $\D \equiv \h - \ta$ between the Hubble horizon and the $\z$-horizon.
As long as $h>0$ and $\tau < 0$,~(\ref{hub}) implies
\be
\D' = \e > 0\,.
\label{hubexD}
\ee
Thus, when $\h$ is inside the $\z$-horizon, corresponding to $\Delta < 0$, the gap between the horizons narrows; when $\h$ is outside the $\z$-horizon, corresponding to $\Delta >0$, the gap between the horizons widens.  Meanwhile, in this regime~(\ref{boxep}) becomes
\be
(\log \sqrt{\e})' = \ta^{-1} - (\ta + \D)^{-1}\,. 
\label{epexD}
\ee
Unlike Case $i)$, the evolution of $\e$ is no longer necessarily monotonic: when $\h$ is inside the $\z$-horizon, corresponding to $\Delta < 0$, $\epsilon$ decreases; when $\h$ is outside the $\z$-horizon, corresponding to $\Delta >0$, 
$\epsilon$ increases.

It is straightforward to show that all solutions in this case must have emerged from a big bang singularity (where $|h^{-1}| = 0$) a finite time $\tau_{\rm i} < \tau_{\rm fid}$ in the past.  In particular, $|h^{-1}|$ is guaranteed to lie within the $\zeta$-horizon at early times.  Whether this remains the case subsequently depends on initial conditions. Qualitatively, if $|h^{-1}|$ remains within the $\zeta$-horizon, the solution describes a universe that expands forever. This case, which includes the inflationary solution, is described below. If $|h^{-1}|$ instead exits the $\zeta$-horizon, the expansion inevitably comes to a halt at $\tau=0$, and the universe enters a collapsing phase which terminates in a big crunch singularity. This {\it apex} solution is described in Case $iii)$.

Let us now focus on the case where $|h^{-1}|$ stays inside the $\zeta$-horizon, {\it i.e.} $\Delta < 0$. Since $|h^{-1}| < |\tau| < |\tau_{\rm fid}|$ for $\tau_{\rm fid}<\tau < 0$,
$h$ cannot change sign as long as $\tau< 0$, hence $a$ increases monotonically. From the discussion below~(\ref{epexD}), $\epsilon$ shrinks monotonically.
In fact, since $|h^{-1}| < |\tau|$ by assumption, $\h$ must hit zero at some $\t_{\rm f} < 0$. In other words, this case spans a finite time interval $\tau_{\rm i} \leq \tau \leq \tau_{\rm f}$,
during which $a(\tau)$ expands from $0$ to $\infty$, while $\e$ shrinks from $\infty$ to $0$. When $\e=1$, $\h$ reaches a global maximum, and,
as in the contracting case, we can choose this as our fiducial time: $\tau_{\rm fid}\rightarrow \tau_{\rm max} \equiv m^{-1}$, where
$a_{\rm max} = 1$ and $\e_{\rm max} = 1$. The solutions can once again be indexed by $c$ defined in~(\ref{cdef}). 

Unlike the contracting case, $c$ is bounded from above: $|h^{-1}_{\rm max}|$ lies inside the $\zeta$-horizon by assumption, hence $c < 1$.  
For $|h^{-1}|$ to remain within the $\z$-horizon subsequently, we numerically find a tighter bound $c \leq c_0 \approx 0.52$.  As $c$ approaches $c_0$, $\tau_{\rm f}$ 
comes arbitrarily close to 0.

In fact, $c\approx c_0$ is desirable to generate a broad range of modes, since $k_{\rm max}/k_{\rm min} = |\tau_{\rm i}|/|\tau_{\rm f}|$. In this limit, $|h^{-1}|$ grazes the $\zeta$-horizon, corresponding to $\epsilon \ll 1$ and $|\eta| \equiv H^{-1}|{\rm d}\ln\epsilon/{\rm d}t|\ll 1$.
In other words, this case relies on a phase of {\it slow-roll inflation} to generate a broad range of modes. (Because we focus on exact scale invariance,
this is a special case of slow-roll inflation. In particular, at linear order $\epsilon$ and $\eta$ are related in such a way that $n_s - 1 = -2\epsilon - \eta = 0$.) 
The inflationary phase thus generates $N_{\rm inf} \sim \log (1/m \abs{\t_{\rm f}})$ e-folds of scale invariant modes, whereas mode production prior to the onset of the inflationary phase is negligible. Since $|h^{-1}| < |\t|$ throughout, modes exit Hubble before they freeze-out.

\subsection{Apex Branch} In this case $|h^{-1}|$ exits the $\zeta$-horizon at some time $\tau_{\rm exit} < 0$ after the universe emerges from the big bang singularity. Once this happens, there is no turning back --- $\Delta$ becomes positive, and from~(\ref{hubexD}) the gap keeps on growing
for $\tau < 0$.

From the discussion below~(\ref{epexD}), $\e$ attains a local minimum at horizon equality. The exit, defined by $|h^{-1}_{\rm exit}|/|\tau_{\rm exit}| = 1$,
happens only once, so it is a convenient place to set $a_{\rm exit} = 1$. This family of solutions can therefore be indexed by a single parameter,
$\e_{\rm exit} > 0$. 

After horizon equality, the expansion inevitably comes to a halt at $\tau = 0$, at which time (the ``apex") the universe
enters a phase of contraction. The subsequent evolution can be deduced by noting that~(\ref{hub}) and~(\ref{boxep}) are manifestly invariant
under $h\rightarrow -h$, $\tau\rightarrow -\tau$. In other words, evolving forward in time when
$h > 0$ and $\t < 0$ is the same as evolving backwards in time when $h<0$ and $\t>0$.  It follows that $|h^{-1}|$ is guaranteed to reenter
the $\zeta$-horizon, after which it will hit zero at finite $\tau_{\rm f} > 0$, corresponding to a big crunch.

To get a broad range of super-Hubble modes, we need $\epsilon_{\rm exit} \ll 1$ (corresponding to $c\approx c_0$). This leads to a slow-roll inflationary phase, which
occurs as before while $|h^{-1}|$ grazes the $\zeta$-horizon, followed by an {\it expanding adiabatic ekpyrotic} phase~\cite{austin}, during
which $|h^{-1}| \gg |\t|$, $\epsilon\sim 1/\tau^2$ and $a(\tau)$ is slowly expanding. This solution thus includes two distinct phases of appreciable mode production.

The inflationary phase generates $N_{\rm inf} \sim \epsilon_{\rm exit}^{-1}$ e-folds of scale invariant modes.
Rescaling coordinates to set $a = 1$ when $\tau = 0$, outside the $\zeta$-horizon $h^{-1}$ satisfies
\be
m h^{-1} \approx - m^{-1} \t^{-1}  - m(1- \e_{\rm exit}) \t + \e_{\rm exit}^{-1/2}\,.
\label{hapex}
\ee
Substituting in~(\ref{hub}), we see that the ekpyrotic phase with $\epsilon\approx 1/m^2\tau^2$
begins at $\tau_{\rm ek-beg} \approx - m^{-1}\e_{\rm exit}^{-1/2}$. This phase ends when Hubble re-enters
the $\zeta$-horizon, which from~(\ref{hapex}) occurs at $\tau_{\rm ek-end} \approx m^{-1}\sqrt{\e_{\rm exit}}$.

The apex marks the end of mode generation.  For $\tau > 0$, modes begin to
re-enter the $\zeta$-horizon, spoiling their scale invariance. Modes with $k \tau_{\rm ek-end} > 1$ end up {\it not} scale invariant. The adiabatic ekpyrotic phase thus
generates $N_{\rm ek} = \log |\tau_{\rm ek-beg}|/|\tau_{\rm ek-end}| \approx \log \epsilon_{\rm exit}^{-1}$ e-folds
of scale invariant, super-Hubble modes. (Arbitrarily many e-folds can be obtained by
ending this phase near $\tau=0$ while the modes remain within Hubble, but a subsequent phase
would be necessary to push these modes outside Hubble while preserving their spectrum~\cite{austin}.)

\section{Non-gaussianities}

While the two non-inflationary branches which rely on a rapidly-varying $\epsilon(t)$ yield power spectra identical to that of inflation,
the degeneracy with inflation is broken by non-gaussianities. The 3-point amplitude for the contracting adiabatic
ekpyrotic mechanism was calculated in detail in~\cite{adiabaticek2}. The resulting non-gaussianities are strongly scale
dependent and peak on small scales, with the dominant contribution growing as $k^2$. Since the 3-point calculation of~\cite{adiabaticek2}
ignored the time-dependence of the scale factor, to a good approximation the result applies equally well to the contracting or apex case.  For completeness, we reproduce here the salient points of the 3-point amplitude calculation in the contracting case.

To make contact with the results in~\cite{adiabaticek2}, we introduce the parameter $H_0 \equiv -m/c$, where $c$ was defined in~(\ref{cdef}).  To see the physical significance of $H_0$, note that~(\ref{conhub}) implies that during the adiabatic ekpyrotic phase, $-c/2m \leq \t  \leq -2/m c$, $h^{-1}$ is within about a factor of two of its maximum value, $h^{-1}_{\rm max} = H_0^{-1}$.  It follows that $h^{-1}$ is nearly constant and
\be
h^{-1} \sim H_0^{-1}
\ee
until near the very end of the phase.  The parameter $H_0$ is thus the characteristic Hubble parameter during this phase. Furthermore, 
the end points of the contracting adiabatic ekpyrotic phase, $\tau_{{\rm eq} +}$ to $\tau_{{\rm eq} -}$, are given by
\be
\tau_{{\rm eq} +} \approx \frac{1}{2H_0}\,; \qquad \tau_{{\rm eq} -} \approx \frac{1}{c^2 H_0}\,.
\ee
Thus the long-wavelength cutoff for our calculations is $\tau_{{\rm eq} +}\sim H_0^{-1}$.
The short-wavelength cutoff is $\tau_{{\rm eq} -}$, which is suppressed by a factor of $1/c^2 \ll 1$ relative to the long-wavelength scale.

The cubic action for $\zeta$ corresponding to
a canonical scalar field with unit sound speed is given by, up to a field redefinition,~\cite{maldacena}
\begin{eqnarray} 
\label{action3}
S_{3}&\simeq &\int {\rm d}t {\rm d}^3x \bigg\{
\epsilon^2\zeta\dot{\zeta}^2 + \epsilon^2\zeta(\vec{\nabla}\zeta)^2-
2 \epsilon\dot{\zeta}\vec{\nabla}
\zeta\cdot \vec{\nabla} \chi \nonumber \\ &+&
\frac{\epsilon}{2}\dot{\eta}\zeta^2\dot{\zeta}
+\frac{\epsilon}{2}\vec{\nabla}\zeta\cdot\vec{\nabla}
\chi \nabla^2 \chi +\frac{\epsilon}{4}\nabla^2\zeta(\vec{\nabla}
\chi)^2 \bigg \},
\end{eqnarray}
where spatial derivatives are contracted with the Euclidean metric $\delta_{ij}$, 
and $\chi$ is defined as $\nabla^2 \chi = \epsilon\dot{\zeta}$. Moreover, following~\cite{adiabaticek2} we have ignored the time-dependence of
the scale factor and set $a\simeq 1$.  At first order in perturbation theory and in the interaction picture, the three-point function is
\begin{eqnarray} 
\nonumber
\langle
& & \zeta(t,\textbf{k}_1)\zeta(t,\textbf{k}_2)\zeta(t,\textbf{k}_3)\rangle = \\
& & -i\int^{t_0}_{-\infty}{\rm d}t^{\prime}\langle[
\zeta(t,\textbf{k}_1)\zeta(t,\textbf{k}_2)\zeta(t,\textbf{k}_3),H_{\rm int}(t^{\prime})]\rangle \,,
 \label{interaction}
\end{eqnarray}
where $H_{\rm int} = - L_{3}$, up to interactions that are higher-order in the number of fields, and $t_0$ is chosen to be sufficiently late that all
modes of interest have frozen out by this time. A natural choice in our case is 
\be
t_0 = \tau_{{\rm eq} -} \approx  \frac{1}{c^2 H_0}\,.
\label{t0}
\ee
As usual it is convenient to express the three-point function by factoring out appropriate powers of the power spectrum and defining an amplitude ${\cal A}$ as follows
\begin{equation}
\langle \zeta(\textbf{k}_1)\zeta(\textbf{k}_2)\zeta(\textbf{k}_3)\rangle = (2\pi)^7 
\delta^3\left(\sum \kk_i\right) P_\zeta^{\;2} \frac{{\cal A}}{\prod_j k_j^3}\,,
\label{Adef}
\end{equation}
where $P_\zeta \equiv k^3|\zeta_k|^2/2\pi^2$ is the power spectrum for the curvature perturbation.

The three-point function receives contributions from each interaction term in~(\ref{action3}). The dominant contributions, as shown in~\cite{adiabaticek2},
are the last two terms in~(\ref{action3}), both of which are ${\cal O}(\epsilon^3)$. The next-to-leading contribution is the $\dot{\eta}$ term. We briefly review the calculation
of these two contributions and refer the reader to~\cite{adiabaticek2} for further details.

The $\epsilon^3$ terms give the combined interaction Hamiltonian
\be
H_{\rm int} =  -\frac{\epsilon^3}{4}\int {\rm d}^3x\left(\nabla^2\zeta\frac{\vec{\nabla}}{\nabla^2}\dot{\zeta}\frac{\vec{\nabla}}{\nabla^2}\dot{\zeta}+2\dot{\zeta}\vec{\nabla}\zeta \frac{\vec{\nabla}}{\nabla^2}\dot{\zeta}\right)\,.
\ee
Applying the canonical commutation relations, the three-point correlation function~(\ref{interaction}) in this case reduces to
\begin{multline}
\langle \zeta(\textbf{k}_1)\zeta(\textbf{k}_2)\zeta(\textbf{k}_3)\rangle_{\epsilon^3}\, =\,
i (2 \pi)^3 \delta^3\left(\sum \kk_i\right) \prod \zeta_{k_i}(0) \\ 
\nonumber
\times  \int_{-\infty+i\varepsilon}^{t_0} {\rm d} t \Bigg\{ \frac{\epsilon^3}{4} \left(\frac{k_1^2}{k_2^2}\zeta_{k_1}^*(t)\,\frac{{\rm d} \zeta_{k_2}^*(t)}{{\rm d} t} + 2\frac{{\rm d} \zeta_{k_1}^*(t)}{{\rm d} t}\zeta_{k_2}^*(t) \right)\\
\frac{\vec{k}_2\cdot \vec{k}_3}{k_3^2}\frac{{\rm d} \zeta_{k_3}^*(t)}{{\rm d} t} + {\rm perm.} + {\rm c.c.}\Bigg\} \,,
\label{eps33pt}
\end{multline}
where the small imaginary part at $t\rightarrow -\infty$ projects onto the adiabatic vacuum state. 
Using the mode functions~(\ref{vk}) and substituting $\epsilon(t)\simeq 1/m^2t^2$, it is easy to show that the integrand
is a total derivative:
\begin{equation}
\int^{t_0} {\rm d}t \, \frac{3-iKt}{t^4}e^{iKt} = - \int^{t_0} {\rm d}t \, \frac{{\rm d}}{{\rm d}t}\left(\frac{e^{iKt}}{t^3}\right) = -c^6 H_0^3\,,
\end{equation}
where in the last step we have used~(\ref{t0}) and taken the long
wavelength limit $K \ll |\tau_{{\rm eq} -}|^{-1} \approx c^2 |H_0|$, which is appropriate for the modes of interest. Putting everything together, the three-point amplitude is~\cite{adiabaticek2}
\begin{equation}
{\cal A}_{\epsilon^3} = \frac{K^2}{32 H_0^2}\left(\sum_ik_i^3 - \sum_{i\neq j}k_ik_j^2+2k_1k_2k_3\right)\,.
\label{eps3}
\end{equation}
As claimed, the 3-point amplitude is strongly scale dependent and peaks on small scales.

The next-to-leading order contribution comes from the $\dot{\eta}$ vertex in~(\ref{action3}):
\begin{equation}
H_{\rm int} = -\int {\rm d}^3x \,\frac{1}{2}\epsilon\dot{\eta} \zeta^2\dot{\zeta}  \,.
\end{equation}
Using the fact that $\dot{\eta} \simeq -2m^{-1}t^{-2}$,  the three-point amplitude is given by, in the long wavelength ($K \ll c^2 |H_0|$) limit,~\cite{adiabaticek2}
\be
{\cal A}_{\dot{\eta}} = - \frac{\pi}{8}\frac{K}{H_0} \left( \frac{K}{2}\sum_ik_i^2 - \sum_{i\neq j}k_ik_j^2 +k_1k_2k_3\right)\,.
\label{Adoteta}
\ee
This contribution scales as $K/|H_0|$ and is therefore subdominant relative to~(\ref{eps3}) on scales $K\gsim |H_0|$.
All other contributions to the three-point amplitude are suppressed by $1/c^2 \ll 1$ relative to~(\ref{eps3}).

Following standard conventions, the three-point amplitude translates into a value for $f_{\rm NL}^{\rm equil.}$, defined at the
equilateral configuration:
\be
f_{\rm NL}^{\rm equil.} \equiv 30\frac{{\cal A}_{k_i=K/3}}{K^3}  \simeq - \frac{5}{144}\frac{K^2}{H_0^2}\,.
\label{fNLfinal}
\ee
Unlike the power spectrum, the three-point function is thus strongly scale dependent: $f_{\rm NL}^{\rm equil.}$ is $\lsim\; {\cal O}(1)$
on the largest scales ($K \sim |H_0|$) and grows as $K^2$. Hence, as advocated, the degeneracy with inflation is badly broken by non-Gaussianities.

Since the perturbative parameter is $f_{\rm NL}\zeta$, with $\zeta\sim 10^{-5}$, perturbation theory breaks down for $K \;\gsim\; 10^{5/2} |H_0|$. 
In fact, on even smaller scales, $K \gsim 10^5 |H_0|$, quantum corrections dominate the classical answer, signaling strong coupling~\cite{adiabaticek2}.
As argued in~\cite{adiabaticek2}, however, these pathologies can be circumvented by modifying the $\epsilon \sim 1/t^2$ behavior before the
dangerous modes are generated. In that case, the power spectrum for $\zeta$ tilts strongly to the red and then flattens out at an exponentially
smaller amplitude with an acceptable non-gaussianity ($f_{\rm NL} \zeta \ll 1$) throughout. This results in a finite window ($|H_0|  \;\lsim \; K \;\lsim\; 10^5 |H_0|$)
of scale invariant modes, which is sufficient to account for large scale structure and microwave background observations.

\section{Concluding Remarks}
We have uncovered three distinct cosmological phases that yield a broad range of
scale invariant modes: inflationary expansion, adiabatic ekpyrotic contraction~\cite{adiabaticek}, and adiabatic
ekpyrotic expansion~\cite{austin}. All three phases generate identical power spectra for $\zeta$, and each is an
attractor background.

The degeneracy is broken at the 3-point level. The rapidly-varying equation of state characteristic of adiabatic ekpyrotic phases
results in strongly scale-dependent non-gaussianities~\cite{adiabaticek2}.  Our results imply that inflation is the unique
single-field mechanism with unit sound speed capable of generating a broad range of scale invariant and gaussian modes.

Forthcoming work~\cite{austin} will extend the analysis to include a general sound speed
$c_s(\tau)$, the other degree of freedom at our disposal~\cite{piazza}. 


{\it Acknowledgments:}  We thank D.~Baumann, A.~Joyce, L.~Leblond, J.-L.~Lehners, S.~Shandera, P.~J.~Steinhardt and M.~Zaldarriaga for helpful discussions.
This work is supported in part by the US Department of Energy (DE-AC02-76-ER-03071) and the Alfred P. Sloan Foundation.


\begin{thebibliography}{99}

\bibitem{zeta}
 J.~M.~Bardeen, P.~J.~Steinhardt and M.~S.~Turner,
  Phys.\ Rev.\  D {\bf 28}, 679 (1983);
 V.~F.~Mukhanov, H.~A.~Feldman and R.~H.~Brandenberger,
  Phys.\ Rept.\  {\bf 215}, 203 (1992).

\bibitem{separateuniverse}
  D.~S.~Salopek and J.~R.~Bond,
  Phys.\ Rev.\  D {\bf 43}, 1005 (1991).

\bibitem{weinbergzeta}
  S.~Weinberg,
  Phys.\ Rev.\  D {\bf 67}, 123504 (2003).

\bibitem{dust}
D.~Wands, Phys.\ Rev.\  D {\bf 60}, 023507 (1999);
F.~Finelli and R.~Brandenberger,  Phys.\ Rev.\  D {\bf 65}, 103522 (2002).

\bibitem{oldek1}
  J.~Khoury, B.~A.~Ovrut, P.~J.~Steinhardt and N.~Turok, Phys.\ Rev.\  D {\bf 64}, 123522 (2001);
   Phys.\ Rev.\  D {\bf 66}, 046005 (2002).

\bibitem{oldek2}
  R.~Y.~Donagi, J.~Khoury, B.~A.~Ovrut, P.~J.~Steinhardt and N.~Turok,
  JHEP {\bf 0111}, 041 (2001).

\bibitem{oldek3}
  J.~Khoury, B.~A.~Ovrut, N.~Seiberg, P.~J.~Steinhardt and N.~Turok,
  Phys.\ Rev.\  D {\bf 65}, 086007 (2002).

\bibitem{oldek4}
  A.~J.~Tolley, N.~Turok and P.~J.~Steinhardt,
  Phys.\ Rev.\  D {\bf 69}, 106005 (2004).

\bibitem{oldek5}
  J.~Khoury, P.~J.~Steinhardt and N.~Turok,
  Phys.\ Rev.\ Lett.\  {\bf 91}, 161301 (2003).
  
 \bibitem{oldek6}
  J.~Khoury, P.~J.~Steinhardt and N.~Turok,
  Phys.\ Rev.\ Lett.\  {\bf 92}, 031302 (2004).
  

\bibitem{gratton}
  S.~Gratton, J.~Khoury, P.~J.~Steinhardt and N.~Turok,
  Phys.\ Rev.\  D {\bf 69}, 103505 (2004).
  
\bibitem{paoloniczal}
  P.~Creminelli, A.~Nicolis and M.~Zaldarriaga, Phys.\ Rev.\  D {\bf 71}, 063505 (2005).
 
\bibitem{robertek}
  R.~Brandenberger and F.~Finelli, JHEP {\bf 0111}, 056 (2001);
  D.~H.~Lyth, Phys.\ Lett.\  B {\bf 524}, 1 (2002).

\bibitem{newek}
    E.~I.~Buchbinder, J.~Khoury and B.~A.~Ovrut,
    Phys.\ Rev.\  D {\bf 76}, 123503 (2007);
     JHEP {\bf 0711}, 076 (2007);
     Phys.\ Rev.\ Lett.\  {\bf 100}, 171302 (2008).
 
 \bibitem{2fieldek}
  J.-L.~Lehners, P.~McFadden, N.~Turok and P.~J.~Steinhardt,
  Phys.\ Rev.\  D {\bf 76}, 103501 (2007);
   F.~Finelli, Phys.\ Lett.\  B {\bf 545}, 1 (2002);
      P.~Creminelli and L.~Senatore,
  JCAP {\bf 0711}, 010 (2007);
    K.~Koyama and D.~Wands,
   JCAP {\bf 0704}, 008 (2007).

\bibitem{adiabaticek}
  J.~Khoury and P.~J.~Steinhardt,
  Phys.\ Rev.\ Lett.\ {\bf 104}, 1301 (2010).
  
\bibitem{linde}
 A.~Linde, V.~Mukhanov and A.~Vikman,
 JCAP {\bf 1002}, 006 (2010).
    
\bibitem{adiabaticek2}
  J.~Khoury and P.~J.~Steinhardt,
  arXiv:1101.3548 [hep-th], accepted for publication in Phys.\ Rev.\  D.

\bibitem{austin}
A.~Joyce and J.~Khoury,
arXiv:1104.4347 [hep-th].

\bibitem{Kinney:2010qa}
  W.~H.~Kinney and A.~M.~Dizgah,
  Phys.\ Rev.\  D {\bf 82}, 083506 (2010).

\bibitem{csm}
  J.~Magueijo and J.~Noller, Phys.\ Rev.\  D {\bf 81}, 043509 (2010).
  
\bibitem{csk}
  D.~Bessada, W.~H.~Kinney, D.~Stojkovic and J.~Wang,
  Phys.\ Rev.\  D {\bf 81}, 043510 (2010).
  
\bibitem{csfedo}
  J.~Magueijo, J.~Noller and F.~Piazza,
   Phys.\ Rev.\  D {\bf 82}, 043521 (2010). 

\bibitem{csearlier}
   C.~Armendariz-Picon and E.~A.~Lim,
   JCAP {\bf 0312}, 002 (2003);
     C.~Armendariz-Picon,
  JCAP {\bf 0610}, 010 (2006);
  Y.~S.~Piao,
  Phys.\ Rev.\  D {\bf 75}, 063517 (2007).

\bibitem{piazza}
   J.~Khoury and F.~Piazza,
  JCAP {\bf 0907}, 026 (2009).

\bibitem{Erick}
J.K. Erickson, D.H. Wesley, P.J. Steinhardt and N. Turok,
{\it Phys. Rev. D}{\bf 69}, 063514 (2004).

\bibitem{ghostNEC}
  P.~Creminelli, M.~A.~Luty, A.~Nicolis and L.~Senatore,
  JHEP {\bf 0612}, 080 (2006);
  P.~Creminelli, A.~Nicolis and E.~Trincherini,
  JCAP {\bf 1011}, 021 (2010).

\bibitem{boyle}
  L.~A.~Boyle, P.~J.~Steinhardt and N.~Turok,
  Phys.\ Rev.\  D {\bf 70}, 023504 (2004).
  
 \bibitem{maldacena}
  J.~M.~Maldacena,
  JHEP {\bf 0305}, 013 (2003)
  [arXiv:astro-ph/0210603].

 
   

\end{thebibliography}
\end{document}